\documentclass[sigconf]{acmart}
\usepackage{url}
\usepackage{xac}

\AtBeginDocument{%
  \providecommand\BibTeX{{%
    \normalfont B\kern-0.5em{\scshape i\kern-0.25em b}\kern-0.8em\TeX}}}

\copyrightyear{2025}
\acmYear{2025}
\setcopyright{cc}
\setcctype{by}
\acmConference[UIST '25]{The 38th Annual ACM Symposium on User Interface Software and Technology}{September 28-October 1, 2025}{Busan, Republic of Korea}
\acmBooktitle{The 38th Annual ACM Symposium on User Interface Software and Technology (UIST '25), September 28-October 1, 2025, Busan, Republic of Korea}
\acmDOI{10.1145/3746059.3747747}
\acmISBN{979-8-4007-2037-6/2025/09}

\begin{document}

\title[CoSight]{CoSight: Exploring Viewer Contributions to Online Video Accessibility Through Descriptive Commenting}

\author{Ruolin Wang}
\authornote{This work was done when the author was at UCLA HCI Research.}
\affiliation{%
  \institution{Georgia Institution of Technology}
  \city{Atlanta}
  \state{GA}
  \country{United States}
}
\email{violynne@gatech.edu}

\author{Xingyu Bruce Liu}
\affiliation{%
  \institution{UCLA HCI Research}
  \city{Los Angeles}
  \state{CA}
  \country{United States}
}
\email{xingyuliu@ucla.edu}

\author{Biao Wang}
\affiliation{%
  \institution{Cornell University}
  \city{Ithaca}
  \state{NY}
  \country{United States}
}
\email{bw437@cornell.edu}

\author{Wayne Zhang}
\affiliation{%
  \institution{UCLA HCI Research}
  \city{Los Angeles}
  \state{CA}
  \country{United States}
}
\email{waynezhang99@ucla.edu}

\author{Ziqian Liao}
\affiliation{%
  \institution{Harvard University}
  \city{Boston}
  \state{MA}
  \country{United States}
}
\email{zliao@hsph.harvard.edu}

\author{Ziwen Li}
\affiliation{%
  \institution{UCLA HCI Research}
  \city{Los Angeles}
  \state{CA}
  \country{United States}
}
\email{zil105@ucla.edu}

\author{Amy Pavel}
\affiliation{%
  \institution{University of Texas, Austin}
  \city{Austin}
  \state{TX}
  \country{United States}
}
\email{apavel@cs.utexas.edu}

\author{Xiang `Anthony' Chen}
\affiliation{%
  \institution{UCLA HCI Research}
  \city{Los Angeles}
  \state{CA}
  \country{United States}
}
\email{xac@ucla.edu}

\renewcommand{\shortauthors}{Wang et al.}


\begin{abstract}
The rapid growth of online video content has outpaced efforts to make visual information accessible to blind and low vision (BLV) audiences. While professional Audio Description (AD) remains the gold standard, it is costly and difficult to scale across the vast volume of online media. In this work, we explore a complementary approach to broaden participation in video accessibility: engaging everyday video viewers at their watching and commenting time. We introduce \textit{CoSight}, a Chrome extension that augments YouTube with lightweight, in-situ nudges to support descriptive commenting. Drawing from Fogg’s Behavior Model, CoSight provides visual indicators of accessibility gaps, pop-up hints for what to describe, reminders to clarify vague comments, and related captions and comments as references. In an exploratory study with 48 sighted users, CoSight helped integrate accessibility contribution into natural viewing and commenting practices, resulting in 89\% of comments including grounded visual descriptions. Follow-up interviews with four BLV viewers and four professional AD writers suggest that while such comments do not match the rigor of professional AD, they can offer complementary value by conveying visual context and emotional nuance for understanding the videos.
\end{abstract}



\begin{teaserfigure}
  \includegraphics[width=\textwidth]{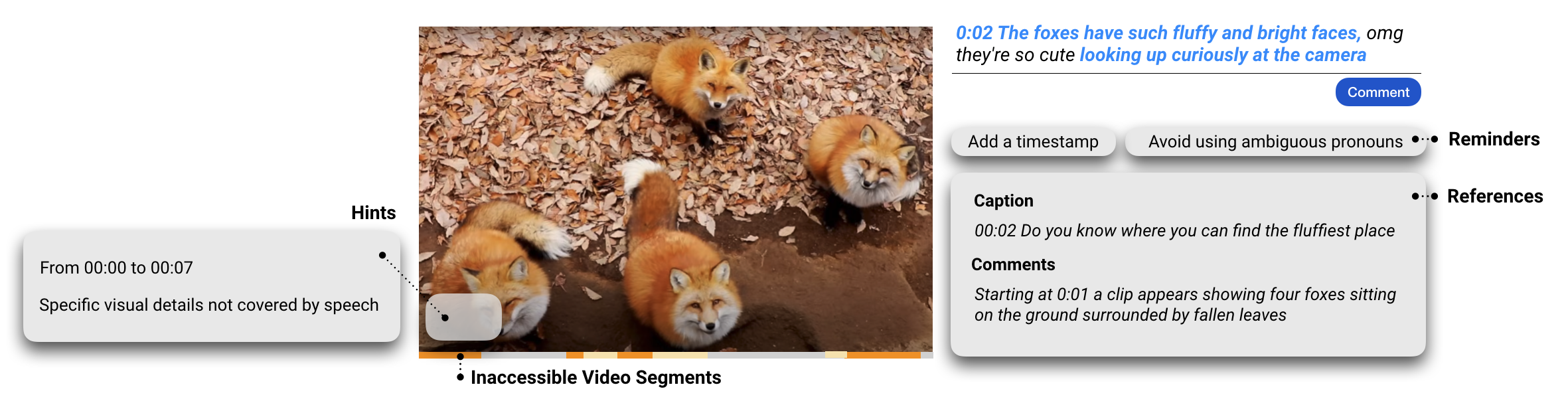}
  \caption{
  CoSight, a Chrome extension developed as a design probe to explore how lightweight interface nudges might encourage accessibility contributions from sighted video viewers when watching and commenting. The prototype augments YouTube video pages with features inspired by Fogg’s Behavior Model~\cite{fogg2009behavior}, including color labels to highlight accessibility gaps (sparks), hints and references to guide contributions (facilitators), and reminders at key moments (signals).
  }
  \Description{\xx}
  \label{fig:teaser}
\end{teaserfigure}

\maketitle
\section{Introduction}
The rapid growth in popularity of online videos has widened the accessibility gap for blind and low vision (BLV) audiences who lack access to the visual content in videos~\cite{liu2021makes}.
Professional Audio Description (AD) --- created by trained describers to provide rich, structured narration --- remains the gold standard for making videos accessible.
However, producing professional AD is resource-intensive~\cite{mikul2010audio}, making it difficult to scale across the immense and growing volume of online content. For example, over 500 hours of new content are uploaded to YouTube every minute\footnote{\url{https://www.globalmediainsight.com/blog/youtube-users-statistics/}} --- a pace that highlights the scalability challenge.
To support and expand the accessibility ecosystem alongside professional AD, prior research has introduced methods to broaden participation in description efforts.
These include engaging novice but dedicated describers such as volunteers and video content creators with specialized tools for writing high-quality descriptions efficiently~\cite{liu2022crossa11y, pavel2020rescribe, yuksel2020human}. Another line of research has focused on developing AI models for automatically generating audio descriptions~\cite{wang2021toward, 10656902}, aiming to provide descriptions at scale. Yet these automated approaches still reveal important limitations, including the absence of fine details, contextual or story-centric elements, and a narration tone that aligns with the video’s theme --- factors that BLV viewers find valuable and important for an engaging experience.

In this work, we explore a new and complementary perspective in the ongoing effort to broaden participation in video accessibility. Rather than relying solely on professional describers, content creators, volunteers, or automated systems, we investigate how casual viewers --- a vastly larger and currently untapped population  --- might contribute to accessibility through comments integrated into their natural viewing experience. Although these viewers typically lack formal training in accessibility or experience with AD, their responses to visual content can offer vivid audience perspectives, capturing emotional cues, contextual nuances, and socially engaging moments. Such comments --- though not a substitute for professional AD --- may help convey visual context in online videos where AD is unavailable.

Prior work in other domains such as online product images~\cite{10.1145/3411764.3445547} and comics~\cite{10.1145/3491102.3502081} has shown that people can unknowingly contribute to accessibility by including visual details in reviews and comments, though the effectiveness of such contributions depends on the clarity and specificity of those details. This challenge also arises in online videos, where comments referencing visuals often lack sufficient context~\cite{10.1145/3290605.3300719}. For instance, a comment like \textit{``2:50 this kid is soooo funny xD''} only expresses an emotional reaction but fails to describe the visual moment itself.
Tapping into the accessibility potential of viewer comments thus requires guiding users toward clearer visual grounding and contextual clarity.

To explore this opportunity, we first conducted an online survey to understand internet users' attitudes toward contributing to the accessibility of different kinds of visual media they consume. Our findings revealed three key factors affecting engagement: \textit{mindset of information-sharing}, \textit{sense of responsibility}, and \textit{balancing efforts and perceived value}. While the first two are shaped by individual contexts and preferences, the third highlights a practical design opportunity: reducing the required effort and clearly communicating the value of contributions can help broaden participation.
Building on these insights and drawing inspiration from the principles of professional audio description, we identified low-barrier actions that align with viewers' natural video-watching and commenting flow, making them approachable for casual contributors. Specifically, we considered two potential ways viewers might contribute to video accessibility through comments: \one{make their comment more accessible}; or \two{write a description for an inaccessible segment}. 
The first is a more intuitive, lightweight action --- such as clarifying vague references and grounding them in specific visual details --- that aligns closely with what viewers already intend to express in their comments.
The second requires greater effort and a clearer understanding of what and how to describe, involving attempts to convey visual content that is not accessible through the audio track.
By observing how viewers engage with these forms of contribution, we aim to uncover what manageable, opportunistic actions are feasible in practice --- and how such contributions might offer value to BLV audiences.

We drew from Fogg's Behavior Model~\cite{fogg2009behavior} as a conceptual framework to design nudges for accessibility-minded actions, including:
\textit{sparks} to boost motivation by drawing attention to accessibility gaps in the video;
\textit{facilitators} to increase ability by offering guidance on what and how to describe; 
\textit{signals} to remind people to consider contributing to accessibility at watching and commenting time. 
As a design probe, we developed CoSight, a Chrome extension that augments YouTube video pages with interactive features embodying these nudges. These include:
color labels of inaccessible video segments (\textit{sparks}), pop-ups with hints on describing inaccessible segments (\textit{facilitators}), related captions and comments as contextual references (\textit{facilitators}), and real-time reminders to ensure comments are clear and visually anchored (\textit{signals}).

We deployed CoSight in an exploratory study with 48 sighted YouTube users to better understand how viewers engage with accessibility-oriented commenting when lightweight guidance is embedded into their video-watching experience. Participants viewed videos from four genres (popular science, documentary, vlog, and how-to) and could freely decide whether, when, and what to comment when watching such videos.
Across 8 videos, participants contributed 257 comments. A majority of comments included timestamps (94\%) and grounded visual references (89\%), describing visual elements such as scenes, actions, objects, text, and animations not covered by the video’s audio. Notably, 93\% of timestamped comments aligned with segments CoSight identified as potentially inaccessible, though viewers also spontaneously commented on other moments. Among comments with concrete visual details, 25\% conveyed emotional reactions paired with visual clarification, while 75\% focused on describing the visual content itself. 
These findings suggest that viewers can make a range of accessibility-relevant contributions through descriptive comments, from visually grounded personal reactions to more deliberate descriptive efforts. Participants found the interaction intuitive; many --- particularly those without prior accessibility experience --- favored lightweight revisions that grounded their comments in specific visual content.

To examine both the promise and limitations of viewer-authored comments, we conducted follow-up interviews with four BLV individuals who regularly use YouTube and have multi-year experience with AD, as well as four professional AD writers. 
We developed an accessibility mode in CoSight that enabled BLV participants to listen to time-anchored comments read aloud by screen readers while watching videos. BLV participants valued comments that conveyed visual details not captured in audio. While some favored objective descriptions, others welcomed subjective elements when paired with useful visual information, noting that such comments could enrich their understanding with added contexts. AD writers noted that while such descriptive comments could not match with professional AD standards, they can offer complementary value in support BLV engagement with video content and raise accessibility awareness among sighted viewers. We discuss potential limitations and risks of viewer-authored descriptive comments for accessibility --- including subjectivity, bias, and potential for abuse --- and outline future directions to support quality and mitigate harm.

In summary, this paper contributes:

\begin{itemize}
    \item An online survey identifying factors that influence media consumers' attitudes toward contributing to accessibility, with design insights  focused on video viewer
    \item CoSight, a Chrome extension developed as a design probe to support descriptive commenting on YouTube, informed by Fogg’s Behavior Model
    \item An exploratory study involving 48 sighted viewers, 4 BLV users, and 4 professional AD writers, providing empirical insights into the feasibility, limitations, and perceived value of viewer-generated accessibility contributions
\end{itemize}

\section{Related Work}
Our work is informed by prior research on authoring tools for writing descriptions, utilizing human-authored texts to enhance accessibility, and fostering contributions for accessibility and social good.

\subsection{Authoring Tools for Writing Descriptions}
Accessible videos for BLV people require a corresponding audio description, or a narration of the important visual details that cannot be understood from the main soundtrack alone~\cite{acbguidelines, packer2015overview, wcag2}. Authoring audio description is a time-consuming process which requires knowledge of where, what and how to describe.
Hiring professional audio describers to provide high-quality audio is costly and not readily-available~\cite{10.1145/3357236.3395433}, as estimated to cost \$12 to \$75 per video minute and the turnaround time from days to weeks~\cite{talkcsun}.
While attempts have been made to automate the generation of audio descriptions for videos, they have revealed shortcomings such as context mismatch, lack of logistic, and missing detailed information~\cite{wang2021toward, gagnon2009towards, 10.1145/3334480.3382821, 10.1145/3357236.3395433, 10656902}.
Notably, BLV people perceive human-authored text descriptions as more valuable than AI-generated text~\cite{duarte2021nipping, pereira2021suggesting}.

Recent work suggested involving video creators, novices amateurs, and volunteers in the process to offer cost-effective audio description with assistive tools ~\cite{10.1145/1878803.1878833, 10.1145/3357236.3395433, 10.1145/3441852.3471201, 10.1145/3334480.3382821, branje2012livedescribe}.
Kobayashi \etal~\cite{10.1145/1878803.1878833} and Branje \etal~\cite{branje2012livedescribe} reported that it is feasible for amateur describers with little and no experience to generate effective audio descriptions supported by technology.
Natalie \etal~\cite{10.1145/3441852.3471201} developed a collaborative tool to support the sighted novice to author audio description with the commentary feedback from either a sighted or a blind reviewer.
Technical solutions are developed to minimize the time and efforts of authoring descriptions by segmenting scenes~\cite{gagnon2009towards}, detecting visual content~\cite{gagnon2009towards, gagnon2010computer, campos2020cinead}, identifying accessibility issues ~\cite{liu2021makes, 10.1145/3526113.3545703}, fitting human-authored descriptions into timeline~\cite{branje2012livedescribe, pavel2020rescribe}, and synthesizing text-to-speech for playback of the descriptions~\cite{3playmedia, gagnon2010computer, 10.1145/1639642.1639699}.
Beyond video accessibility, prior work also pointed out various ways to improve description quality of people by making guidelines easier to understand with visualizations~\cite{bigham2010webtrax, sato2009s, takagi2003accessibility, peng2021say}, providing visual questions as prompts~\cite{10.1145/2764916}, showing high-quality examples \cite{Kobayashi2018AnES, 10.1145/3415176}, and providing assessment \cite{10.1145/3148148, 10.1145/2791285}.

Unlike prior work that supports motivated novice describers working within structured authoring settings and aiming to follow formal AD formats, we explore how everyday viewers, who may not be aware of accessibility gaps, can contribute to video accessibility in-situ as part of their natural viewing and commenting experience. While not aiming to replicate professional AD, which remains essential for equitable access, we draw from established AD guidelines~\cite{dcmp, adlab, fryer2016introduction} to identify core values such as clarity, narrative relevance, and temporal alignment. These values shaped our design of simple, approachable actions that everyday viewers can begin adopting in casual media contexts.

\subsection{Utilizing Human-Authored Texts to Enhance Accessibility}
Crowdsourcing has been widely applied as an effective solution for tackling accessibility challenges in computing, particularly in providing descriptions to visual media such as videos~\cite{youdescribe}, images~\cite{10.1145/1866029.1866080, 10.1145/3313831.3376728} and GIFs~\cite{10.1145/3491102.3502092}.
For example, YouDescribe~\cite{youdescribe} is a free crowdsourcing platform for YouTube that allows BLV people to submit their requests and supports sighted volunteers to script and record audio descriptions.
Lasecki \etal~\cite{10.1145/2380116.2380122} introduced a new approach in which groups of non-experts collectively caption speech in real-time to help deaf people.
Futhermore, organic crowdsourcing~\cite{10.1145/3479863}, which collects useful data as a byproduct of tasks people are already intrinsically motivated to perform, offers a particularly relevant model for our work. This approach has been applied in accessibility contexts such as collecting sign language data~\cite{10.1145/3411764.3445416}, and highlights how leveraging everyday user behaviors can improve scalability compared to traditional, task-oriented crowdsourcing.

In parallel, a growing body of work has explored repurposing human-authored content such as product reviews~\cite{10.1145/3290605.3300602, 10.1145/3234695.3236337, 10.1145/3411764.3445547} or social media comments~\cite{10.1145/2818048.2820013, 10.1145/3491102.3502081} as an untapped source of descriptive information. 
For example, Revamp~\cite{10.1145/3411764.3445547} extracts descriptive details from customer reviews of Amazon products to provide vivid visual descriptions covering attributes including color, size, shape and logo. Cocomix~\cite{10.1145/3491102.3502081} extracts comments from viewers who are familiar with the story context to provide descriptions for relevant webtoon panels.
These studies demonstrate that such human-authored content can provide vivid and contextual descriptions, but such content is uncommon without being intentionally encouraged or supported.
In the video domain, prior research has highlighted the potential of video comments to enhance learning~\cite{chen2022timeline, sung2017exploring}, facilitate communication~\cite{10.1145/2998181.2998256}, assist BLV users in identifying whether a video might be accessible~\cite{10.1145/3411764.3445233},as well as enabling BLV vloggers to engage with their audiences~\cite{seo2021understanding}.
Yarmand \etal~\cite{10.1145/3290605.3300719} found that visual entities are often referenced in YouTube comments. However, their analysis also revealed a deficiency in within-comment context, with many references only mentioning the entities but lacking descriptive details, which would limit their usefulness in helping BLV people to understand video contents.

This suggests a promising but underexplored opportunity: comments are already a natural part of the video-watching experience and are leveraged by BLV viewers. Rather than discarding this medium due to its current limitations, our work explores how viewer comments might be shaped through lightweight scaffolding to carry richer, descriptive visual information that supports accessibility.

\subsection{Fostering Contributions for Accessibility and Social Good}
We are inspired by how lightweight interventions can change people's behaviors on social platforms and doing good for others.
Prior work has shown that subtle, well-placed cues can guide users toward beneficial actions without requiring significant effort.
For example, Jahanbakhsh \etal~\cite{10.1145/3449092} explored lightweight interventions such as checkboxes at posting time to reduce the sharing of misinformation on social media.
More broadly, HCI Researchers have applied the concept of nudging \cite{thaler2009nudge} in multiple contexts including health, sustainability and privacy by designing systems to introduce subtle changes in the way information are presented with the goal of guiding users towards desired behaviors.
Caranan \etal ~\cite{10.1145/3290605.3300733} found 74 examples of nudging in HCI literature, identified 23 distinct mechanisms of nudging, and grouped them in 6 overall categories including facilitate, confront, deceive, social influence, fear, and reinforce.
Nudging is defined as \textit{``any aspect of the choice architecture that alters people’s behavior in a predictable way without forbidding any option or significantly changing their economic incentive''}. Beyond economic incentives, Researchers have also suggested considering other relevant forms of incentives, such as cognitive costs~\cite{thaler2008nudge}, time or sanctions~\cite{hansen2016definition, hausman2010debate}.
Actions that require more deliberate effort or offer rewards, like monetary incentives or intrinsic motivators such as social recognition or volunteer motivation, are generally excluded from the scope of nudging. However, this additional criterion was not directly included in the original definition of nudge, thus creating room for ambiguity~\cite{congiu2022review}.

Recent research has explored ways to engage social media authors in improving image accessibility at posting time.
Bellscheidt \etal ~\cite{bellscheidt2023building} reviewed literature on alt text authoring interfaces for social media and found that most rely on general prompts without focusing on behavior change. They recommend future tools reduce cognitive effort, use motivational and well-timed triggers, and enhance user accountability by making alt text visible and tied to the user's social presence.
Pereira \etal ~\cite{seixas2024automation} developed SONAAR, a tool that encourages users to create accessible content by suggesting text descriptions for images and providing digital accessibility resources.

While prior work has focused on supporting content creators during the upload process, shaping how accessibility is built into future content, our work shifts attention to the viewing side — engaging with the vast body of media already online, much of which lacks accessibility considerations. We explore how viewers might help bridge these existing gaps by contributing through lightweight actions like commenting as they watch.
To support such contributions, we draw on Fogg’s Behavior Model~\cite{fogg2009behavior}, which outlines triggers — sparks, facilitators, and signals — that can motivate, enable, or remind users to act.
Prior work has shown how viewer participation can be harnessed to foster prosocial behavior~\cite{10.1145/3544548.3580912, 10.1145/3710954} and collaborative learning~\cite{10.1145/3274381}. These efforts focus on social dynamics or knowledge curation where viewers themselves benefit from participation. Our work extends this scope to accessibility, where the efforts of sighted viewers aim to provide value for a different group within the same community — BLV audiences.






\section{Exploring Media Consumers' Role in Accessibility: Insights from a Survey}

\begin{figure*}[t]
  \centering
  \includegraphics[width=1.0\linewidth]{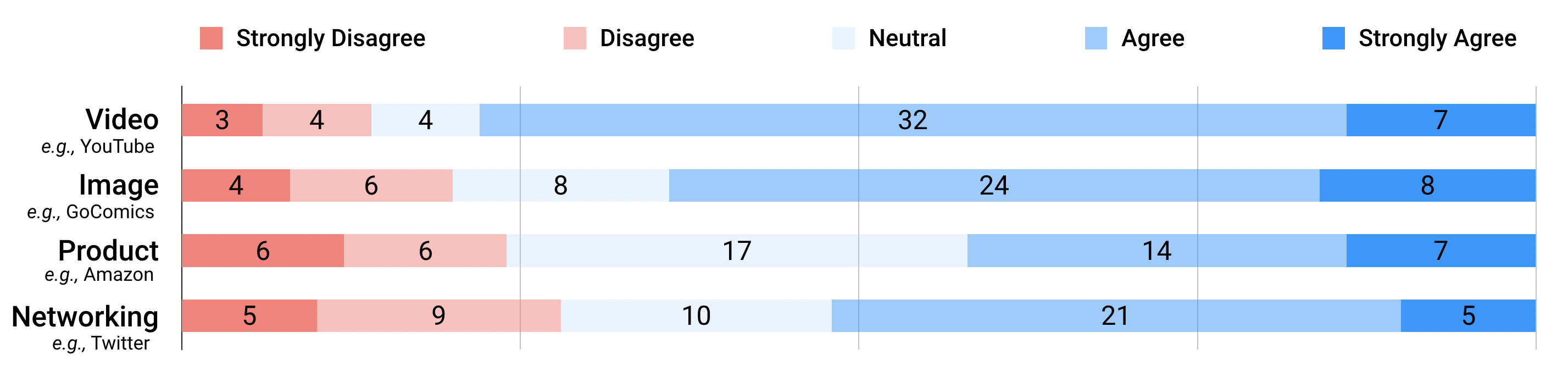}
  \caption{Participants rated their agreements with the statement on a 5-point Likert scale (1 = strongly disagree, 5 = strongly agree): \textit{``I am willing to contribute to media accessibility by additionally providing descriptions of visuals in my comments.''}}
  \label{fig:survey_platform}
\end{figure*}

While prior work has focused on content creators as contributors to accessibility, less is known about the potential role of media consumers --- those who primarily watch, read, or interact with content. To explore whether these everyday users might engage in accessibility-related contributions, and under what conditions, we conducted an online survey with 50 participants.

\subsection{Methods}
We recruited 50 participants through Prolific and compensated each at an hourly rate of \$12 for completing a Google Form-based survey. The participants' ages ranged from 21 to 74 years (M = 41.6, SD = 15.6), with 24 identifying as female and 26 as male, all located in the United States.
Participants rated their agreements with the statement on a 5-point Likert scale (1 = strongly disagree, 5 = strongly agree): \textit{``I am willing to contribute to media accessibility by additionally providing descriptions of visuals in my comments.''} on four categories of social platforms, including video-based platforms (\eg YouTube), image-based platforms (\eg GoComics), product-based platforms (\eg Amazon), networking platforms (\eg Twitter).
Participants were asked to provide concrete reasons for each rating and, at the end of the survey, to also reflect on the factors that influence their attitudes toward contributing on different platforms.
Two researchers followed thematic analysis methods~\cite{gibbs2007thematic} to code the participants' answers and identify the themes.

Although this paper focuses on video-based platforms such as YouTube, we included a range of platform types to provide broader context and minimize potential response bias. Asking only about a single platform may have encouraged participants to give overly favorable responses, particularly if they viewed contributing to accessibility as the “correct” or socially desirable answer. Including multiple platform types would encouraged participants to reflect more critically on the conditions under which they would be willing to contribute.

A limitation of this survey is the relatively small sample size, but it offers valuable insights into video viewers’ willingness and the conditions under which they may contribute to accessibility, shedding light on the feasibility and potential directions for engaging media consumers in accessibility-related efforts.

\subsection{Findings}
Participants reported their willingness to contribute to accessibility (ratings of ``agree'' or ``strongly agree'') across four platform categories, with highest for video-based platforms (40/50), followed by image-based (32/50), networking (26/50), and product-based platforms (21/50), as shown in Figure \ref{fig:survey_platform}.
To better understand these differences, we analyzed participants’ open-ended responses and identified three key factors influencing their attitudes toward contributing to online media accessibility as media consumers: \textit{mindset of information-sharing}, \textit{sense of responsibility}, and \textit{balancing efforts and perceived value}.

\subsubsection{Mindset of Information-Sharing}
Multiple participants (18/50) mentioned their willingness of contributing to accessibility depends on whether their current activities align with a mindset of information sharing. Individual preferences vary, with some participants less inclined to contribute when engaging in entertainment-focused content:
\begin{quote}
\textit{I will definitely be willing to contribute to the accessibility if my goal is about information seeking or discussion. But for entertainment, I assume contributing to that may be overwhelming for me.}
\end{quote}
Others emphasized that even within information-seeking contexts, the urgency of their primary task affected their willingness to contribute:
\begin{quote}
\textit{If I am information seeking, I am usually under a time constraint, and would be less likely to contribute than if I am seeking information for personal use.}
\end{quote}
Participants also highlighted the fragmented nature of attention on fast-paced platforms, where content is consumed quickly and passively—making such environments feel mismatched with thoughtful contribution:
\begin{quote}
\textit{People read and scroll so fast, expect a low mental load. I don’t know if people will be motivated to write things carefully and slowly to contribute to accessibility.}
\end{quote}


\subsubsection{Sense of Responsibility}
Participants’ willingness to contribute was influenced by how connected they felt to a platform’s community and how they understood their role in relation to accessibility.
Many felt more inclined to help in spaces where they were already active or felt part of a community:

\begin{quote}
\textit{``In these online spaces I’m often already sharing information so it’s not going out of my way at all. Plus to me those things have more of a sense of community which makes me feel more like helping.''}
\end{quote}
For media that participants particularly enjoy, such as comics, they reported feeling especially motivated to contribute by sharing that experience with blind and low-vision (BLV) individuals. 
At the same time, some participants emphasized that for media like artworks or product images, the responsibility for ensuring accessibility should lie primarily with the original creators or businesses, rather than the audience:
\begin{quote}
\textit{``I believe that it should fall to the media creator such as the artist providing images and the business owner who sells the product online to do that, and if the audience wants to help then they can but are under no obligation to do so.''}
\end{quote}


\subsubsection{Balancing Effort and Perceived Value}
Participants’ willingness to contribute was shaped by how much effort the task required and whether they felt their input would make a meaningful difference. Many (14/50) said they would be more inclined to help if they understood the impact of their contribution, and if doing so did not disrupt their viewing experience (15/50).
Perceptions of effort varied across media types. Some participants found describing static images easier, while others felt videos were more manageable since the audio already conveys part of the information, leaving them to fill in the gaps:
\begin{quote}
\textit{``I prefer to contribute to videos. because videos already have audio that provides part of the information, so it would be easier to contribute. For comics, photos, it's like you have to do the picture description from zero, which I don't really like.''}
\end{quote}
Perceived value also played a key role. Several participants noted they were more motivated to contribute when they believed their comments could help a larger or more public audience, such as in educational or product-related content:
\begin{quote}
\textit{``It's always about knowing your audience. Sometimes I'm just commenting for friends while other times I post to the masses. The larger the audience, the more reason to be inclusive and be mindful of my comments.''}
\end{quote}
\begin{quote}
\textit{``For casual entertainment I'm less inclined to put in the extra effort to describe a scene. For education I believe extra descriptions may help out everyone.''}
\end{quote}



\subsection{Discussion}
These findings shed light on how everyday media consumers --- who usually watch content for personal reasons --- might be willing to contribute to accessibility as a secondary, helpful act. In what follows, we reflect on how the survey insights shaped the scope of this work and framed our next steps.

\subsubsection{Video as a Promising Context for Accessibility Contributions}
Our findings do not assume that all users are equally ready or responsible for contributing. Instead, they suggest a spectrum of motivations and constraints shaped by personal goals, platform norms, and perceptions of effort and value.

Reflections on mindset of information-sharing indicate that accessibility contributions are more likely when viewers are already inclined to reflect, share, or engage more with content, less so during fragmented, low-effort consumption. Similarly, reflections on sense of responsibility show that viewers are more motivated to help in contexts where they are already active or connected to a community. Many also drew clear boundaries around professionally produced content, such as art or product images, where they believed accessibility should remain the creator’s responsibility.

These insights point to online video as a particularly promising domain for exploration. Watching and commenting on videos are already time-committing and interaction-rich activities, making it easier to build accessibility contributions into natural viewer behavior without asking users to start from scratch.

\subsubsection{Balancing Effort and Perceived Value as an Entry Point for Exploration}
While mindset and responsibility are shaped by long-term platform dynamics that would require large-scale, in-the-wild interventions to meaningfully address, the findings highlight balancing effort and perceived value as a more approachable entry point to explore. 
Participants were more willing to contribute when the required effort felt manageable and the impact of their input was clear. This suggests opportunities to scaffold lightweight, in-situ actions that feel attainable to sighted viewers and offer meaningful value to BLV audiences.

Designing for broader accessibility participation requires navigating a key tension: lowering barriers can invite more engagement, but also risks introducing contributions that may lack clarity or usefulness. Rather than assuming this tension can be addressed by making contribution easier alone, platforms should carefully scaffold participation --- offering simple, timely prompts that align with how viewers naturally watch and comment on videos, while still upholding core values of clarity, narrative relevance, and respect for access needs. By supporting these low-effort but purposeful forms of input, platforms can help everyday viewers, who may not initially see themselves as contributors to accessibility, begin to participate by taking on approachable roles that support accessibility in casual contexts.

\subsubsection{Applying Fogg’s Behavior Model to Support Just-in-Time Contributions}
To guide our design exploration, we applied Fogg’s Behavior Model~\cite{fogg2009behavior} as a conceptual framework to inspire triggers for accessibility contributions, including:
\textit{sparks} to increase motivation of contribution by highlighting the accessibility gaps in the video;
\textit{facilitators} to increase ability of contribution by providing the accessibility knowledge on what and how to describe; 
\textit{signals} to remind people to consider contributing to accessibility at watching and commenting time. 

These categories helped shape how we can deploy a design probe to explore how lightweight, in-situ interface features might support accessibility-relevant contributions from sighted viewers. In the next section, we describe the design and implementation in more detail.

\begin{figure*}[t]
  \centering
  \includegraphics[width=1.0\linewidth]{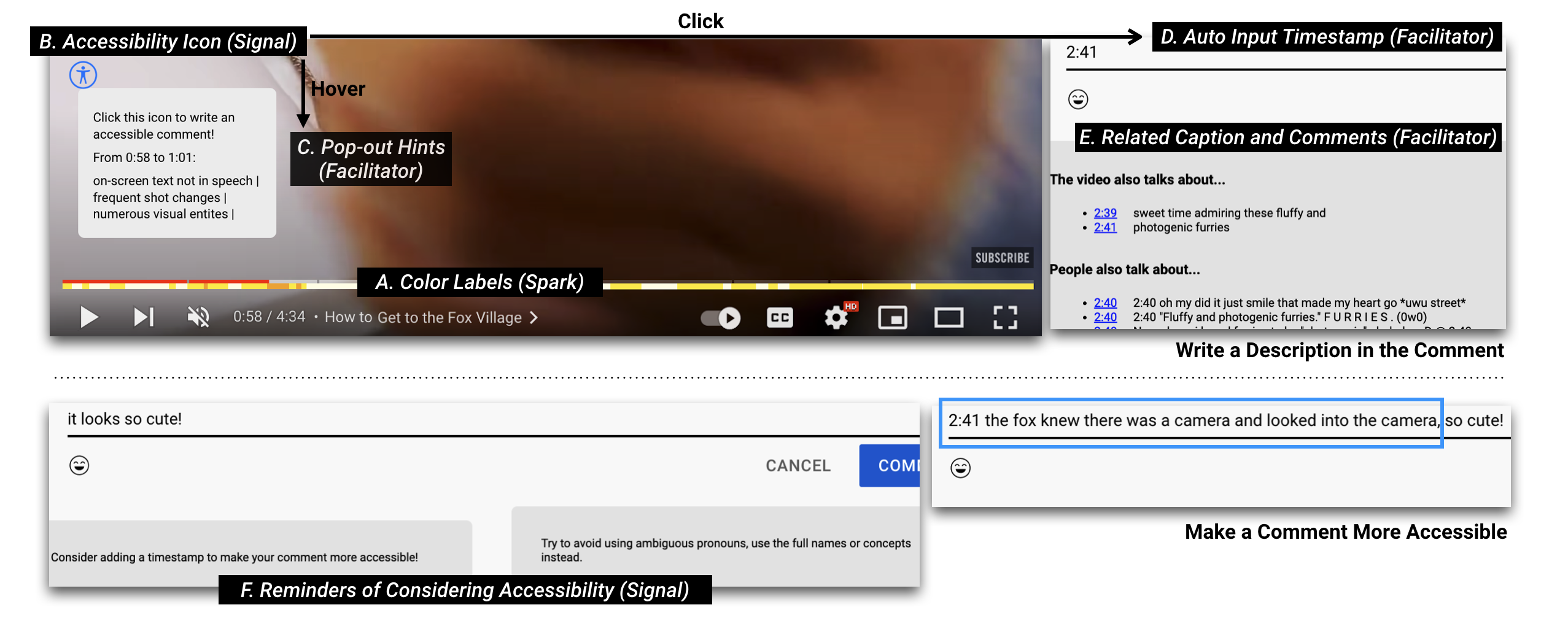}
  \caption{CoSight interaction flow supports two forms of accessibility contributions during video watching and commenting: (1) Make a comment more accessible: While writing a personal comment, related captions/comments appear (facilitator), and real-time reminders prompt users to ground vague references with visual detail (signal). (2) Write a description in the comment: Inaccessible segments are highlighted on the progress bar (spark), with an accessibility icon as a reminder (signal). Hovering reveals a pop-up with visual description hints, and clicking inserts a timestamp into the comment box for writing (facilitator).}
  \label{fig:interaction_flow}
\end{figure*}

\section{CoSight: A Design Probe to Support Descriptive Commenting on YouTube}

To explore how lightweight, just-in-time contributions from video viewers might support accessibility, we developed CoSight, a Chrome extension that augments YouTube’s video page with interactive features. Rather than proposing a fully validated system, CoSight serves as a design probe --- a tool to investigate how interface triggers might encourage sighted viewers to write more descriptive, visually grounded comments as part of their natural interaction with video content.

\subsection{Supporting Descriptive Commenting at Two Levels of Effort}
Building on the previous insight that approachable, in-situ actions can invite broader participation when they feel meaningful and manageable, CoSight supports two levels of contribution designed to reflect core values from professional AD while aligning with natural viewer behavior.

\begin{itemize}

    \item{\textbf{Make a comment more accessible}: Viewers spontaneously writing expressive or reaction-based comments receive lightweight nudges to clarify vague references and anchor their comments to specific visual content. While these contributions are not AD practices, they embody its spirit: helping ensure that viewer-generated content remains comprehensible, relevant, and accessible to BLV audiences. These revisions require minimal effort and are designed to fit naturally into casual commenting behavior.}
    
    \item{\textbf{Write a description in the comment}: For viewers willing to contribute more actively, CoSight streamlines the process of anchoring comments to specific moments and offers guidance on when, what, and how to describe. These contributions resemble audio description in intention though not in structure or quality, as casual viewers are not expected to produce professional-grade narration given their limited training and the constraints of in-the-moment engagement.}

\end{itemize}

These two contribution types are designed to support flexible engagement, allowing viewers to participate in ways that match their comfort and effort levels. To onboard new users and orient them toward accessibility-aware commenting, CoSight provides a dedicated instruction page that introduces the two contribution types and offers simplified guidance grounded in core audio description values. The complete instruction content is included in Appendix.

\subsection{Designing Triggers with Fogg’s Behavior Model}
We drew on Fogg’s Behavior Model~\cite{fogg2009behavior} as a conceptual framework to guide the design of CoSight’s features throughout the video watching and commenting experience.

\textbf{Sparks}: Color labels beneath the video progress bar (Figure~\ref{fig:interaction_flow}A) highlight segments that may be inaccessible to blind and low vision (BLV) viewers, signaling where viewer contributions could be most helpful. More severe gaps are shown in orange, while secondary gaps appear in yellow. These labels increase motivation by indicating how inaccessible is a video and can also help people prioritize their time and efforts on contributing to accessibility.

\textbf{Signals}: As the viewer enters an inaccessible segment, a brief on-screen icon (Figure~\ref{fig:interaction_flow}B) serves as a visual nudge to consider contributing. If the user begins writing a comment and uses vague pronouns (e.g., “this” or “it”) or omits a timestamp, contextual reminders appear (Figure~\ref{fig:interaction_flow}F) to prompt grounding their comment to specific visual details.

\textbf{Facilitators}: Hovering over an accessibility icon that appears during inaccessible segments reveals a pop-up hint (Figure~\ref{fig:interaction_flow}C) describing why the segment may be inaccessible (e.g., ``frequent shot changes''), along with suggestions for how may help (e.g., ``clarify the changing scenes''). 
Clicking the accessibility icon automatically positions the cursor in the comment input box and inserts the current video timestamp (Figure~\ref{fig:interaction_flow}D).
Additional references, such as related captions and previously written comments, are surfaced during comment writing (Figure~\ref{fig:interaction_flow}E) to support users in framing relevant contributions. All accessible comments are collected in a shared list visible on the page, offering a sense of asynchronous collaboration and showing that even small acts can be part of a broader effort.

The interactive features were brainstormed and refined by four researchers with expertise in accessibility and HCI design. Grounded in Fogg’s Behavior Model and iterated based on feasibility and plausibility of viewer interaction on YouTube, these features serve as basic probes—intended to provoke, rather than prescribe, patterns of lightweight accessibility contributions. 

\subsection{Implementation as a Chrome Extension}
CoSight is built with React, TypeScript, MongoDB, and craco npm package. CoSight augmented the Youtube video web pages by injecting the feature modules with Chrome APIs. 

\subsubsection{Identifying and Explaining Inaccessibility}
To identify inaccessible video segments and provide explanations of why a video segment is not accessible, CoSight combined two computational pipelines from prior work on identifying video accessibility issues \cite{10.1145/3411764.3445233, liu2022crossa11y}.CrossA11y \cite{liu2022crossa11y} segments the auditory and visual track of the video and identifies asymmetries between auditory and visual tracks using cross-modal grounding analysis. As a black box method based on the MIL-NCE model~\cite{mle2020end} and the state-of-the-art MultiModal Versatile (MMV) networks~\cite{mmv2020self}, it can not provide concrete reasons on what is inaccessible, but it can rank inaccessibility of video segments by measuring the asymmetries as visual accessibility scores (0.0 - 1.0).
For example, if a video segment contains many visual elements that are not described in the audio, it will receive a low visual accessibility score.
The other work~\cite{10.1145/3411764.3445233} defines video accessibility metrics to explain why one video segment is considered as accessible or inaccessible. For example, if a segment has frequent shot changes, it may be inaccessible and needs more description about the changing scenes.
For people without accessibility knowledge, providing concrete explanations on inaccessible and suggestions on what to describe can help them authoring comments.

CoSight first utilized the CrossA11y pipeline to separate the video into segments and assign each segment an accessibility score, and select the top 5 inaccessible segments by scores, which are labelled as orange color along the progress bar.
CoSight then followed the accessibility metrics to locate second-tier inaccessible segments which are labelled as yellow color along the progress bar, and provide hints for describing the inaccessible segments. 
A full list of explanations provided to users are shown in Appendix.

One limitation of the current implementation is the time cost, which requires around 2 mins processing for one 5-min video. The computational pipelines can be replaced by future available advanced techniques to support real-time processing of any YouTube video, while not the focus of this work.

\subsubsection{Analyzing and Storing Comments}
CoSight used NPM package \textit{keyword-extractor}\footnote{https://www.npmjs.com/package/keyword-extractor} to extract keywords from each caption sentence, as well as from each user comment. These keywords, along with timestamp information, are leveraged to suggest related captions and comments in response to user input.
To select comments for inclusion in the accessible comment list, we filtered out comments that lacked timestamps and had low semantic textual similarity with the corresponding captions, using the SentenceTransformers model from Hugging Face API \footnote{https://huggingface.co/tasks/sentence-similarity}. All the data objects are stored in the Mongo database.

\subsubsection{User Onboarding and Guidance}
Users can register for a CoSight account, and upon logging in, access an onboarding interface via the Chrome menu (Extension - Options). The interface includes three tab pages: Intro, with instructions on why and how comments can support BLV users (as shown in Appendix); Manual, describing CoSight’s interactive features; and Profile, which lists the user’s historic submitted comments.
While basic in form, this implementation provides sufficient scaffolding to support exploratory use and reflection on viewer contributions in our study.

\section{Exploring Viewer Contributions with CoSight: A Two-Stage Study}
To investigate how sighted viewers might engage with accessibility contributions during natural video interactions, we used CoSight as a design probe in a two-stage exploratory study. In the first stage, we recruited 48 sighted YouTube users to watch videos using CoSight and contribute comments. We analyzed their commenting behaviors to understand what kinds of accessibility-relevant content they produced and when. In the second stage, we gathered feedback from four BLV YouTube users and four professional AD writers to examine how these comments were perceived in terms of accessibility value. Together, these two stages helped us explore the feasibility, limitations, and perceived value of lightweight, viewer-generated contributions to online video accessibility.

\subsection{Methods}
\begin{figure*}[h]
  \centering
  \includegraphics[width=1.0\linewidth]{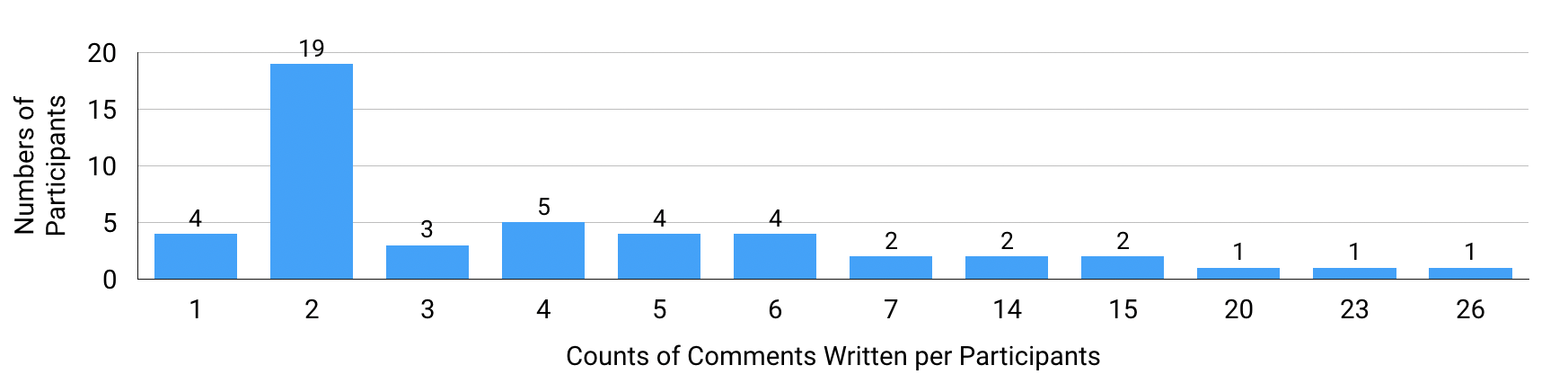}
  \caption{The histogram of counts of comments written per participant. Each participant watched two videos.}
  \label{fig:histogram}
\end{figure*}
\subsubsection{Sighted Viewers Watching Videos with CoSight}
We recruited 48 participants (Female: 23, Male: 25, average age: 37) from Prolific who used YouTube at least once a month. We compensated each participant \$12 for joining the study. Participants were informed that their compensation did not depend on the quantity or quality of their comments, and that commenting was entirely voluntary. Participants downloaded the zip file of our extension demo and installed to Chrome.
They first read instructions on how to write comments for BLV video viewers using the extension. Then, they watched and optionally described two randomly selected videos. Finally, users filled in an online questionnaire to rate the usability of CoSight and answer open-end questions on reflecting helpful features and giving feedback on future improvements. 
Eight videos, covering four categories including \textit{popular science}, \textit{documentary}, \textit{vlog}, and \textit{how-to}, were randomly selected from YouDescribe, a platform where BLV people can request audio descriptions for YouTube videos. Time length of videos range from 2 - 10 mins. 
Two researchers followed the thematic analysis methods~\cite{gibbs2007thematic} to open-code the participant-written video comments and the subjective feedback independently. Then, the researchers reviewed each other's codes and discussed the codes to reach a consensus.


\subsubsection{Interviews with BLV Viewers and AD Writers}
To demostrate for BLV people, we developed an accessibility mode of CoSight interface.
As the BLV viewer plays a YouTube video, a beep notification sound will be delivered at the end of a video segment when there are available comments for this segment.
The video will be paused automatically as default, while user can also change the setting as not pausing automatically. 
BLV user can then press Shift to navigate through and read out the available comments, and press Space to exit the comments reading and resume video playing, which is aligned with the original shortcuts of YouTube.
We followed Web Accessibility Initiative - Accessible Rich Internet Applications (WAI-ARIA) standards to make sure web components of CoSight, such as the list of accessible comments, have proper attributes and keyboard interactions to be compatible with screen readers.

BLV participants (U1-U4) were invited from an online discussion group of Audio Description, with demographic information is shown in in Table \ref{tab:demo}.
The participants walked through the interface of CoSight with the experimenter and played four videos, one in each category. 
Then the experimenter asked participants semi-structured interview questions in two themes: the quality of the comments, and their concerns and expectations on future improvements.
AD writers (W1-W4) were invited from a training program of professional AD writers. 
AD writers walked through the interface of CoSight with the experimenter, tried out the features for sighted viewers when watching two videos of different categories, and gave feedback on the quality of representative collected comments.
All participants were compensated \$20 for the one-hour demonstration and interview.
With user consent, the process conducted on Zoom were recorded and transcribed. One researcher followed the thematic analysis methods~\cite{gibbs2007thematic} to open-code the recorded data and the other researcher reviewed the codes. The researchers met to discuss the themes and reach consensus.

\begin{figure*}[h]
  \centering
  \includegraphics[width=1.0\linewidth]{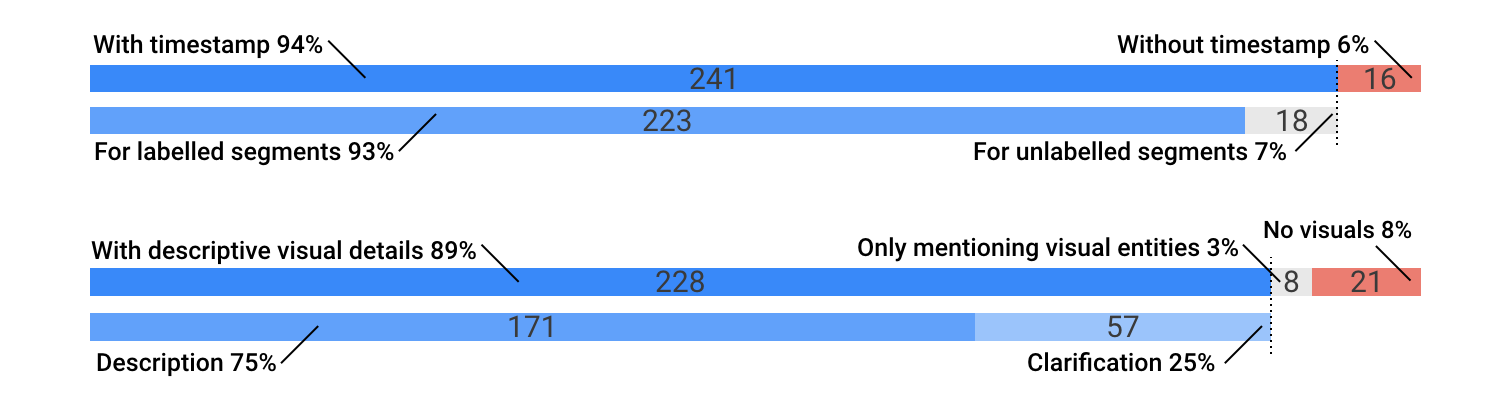}
  \caption{Analysis on the content of 257 comments showed that participants using CoSight included timestamps and descriptive visual details in their comments. Participants wrote comments to describe labeled inaccessible segments, and also clarified visual referents when writing comments to share opinions.}
  \label{fig:feature_expressiveness}
\end{figure*}

\subsection{Analyzing the Content of Comments}
Our analysis demonstrated that participants using CoSight included timestamps and concrete visual details in their comments.
In total, participants wrote 257 comments across 8 videos, ranging in length from 13 to 438 characters (average: 93 characters).
Each participant watched 2 videos but not every participant wrote comments for every video.
4 participants wrote one comment for one out of two videos they watched.
19 participants wrote two comments, one comment for each video they watched.
7 participants wrote more than five comments per video; two of them reported having a friend or family member who is blind.
The histogram of counts of comments written per participant is shown as figure~\ref{fig:histogram}.


\subsubsection{When Did Participants Wrote a Comment?}
As shown in Figure~\ref{fig:feature_expressiveness}, 94\% (241/257) of comments contained a timestamp.
Among all participants, only 4 participants wrote comments without a timestamp. 
2 of them reported that they intended to add timestamps but encountered a technical issue of extension not responding accidentally. 
The remaining comments without a timestamp were general opinions on the whole video, such as \code{I like the detail he is explaining in regards to the nervous system.}\footnote{\url{https://www.youtube.com/watch?v=qPix_X-9t7E}}
93\% (223/241) of comments with a timestamp aligned with the inaccessible segments labelled by CoSight, while the inaccessible segments at the end of a video were likely to be neglected.
For example, at the end of an educational video, there are many links and thumbnails of other related contents not mentioned in the narrations, and labelled as an inaccessible segment. But no participants chose to write a comment for this.

Participants still wrote comments for unlabeled segments when they found visual content necessary to describe, acknowledging that automated methods may not identify every inaccessible segment.
For example, when the narrations introducing the fox village with the scenes walking around, a participant provided an additional visual detail even it was not labelled as inaccessible: 
\code{01:25 a close up on a statue of a red fox}.\footnote{\url{https://www.youtube.com/watch?v=nrSdiqp0Dpk}}
Some participants wrote comments when the segment was accessible, but they only wanted to share an opinion or feeling.
For example, when watching the scene of a lot of cats,
a participant wrote \code{0:19 I would love to visit the cat island!} without providing any additional visual details.


\subsubsection{What Did Participants Describe in Comments?}
As shown in Figure~\ref{fig:feature_expressiveness}, 
89\% (228/257) of the total comments contain concrete visual details not covered by the speech of corresponding video segment, 3\% (8/257) of the total comments mentioned visual entities but didn't covered any concrete descriptive information, 8\% (21/257) of the total comments purely expressed participants' opinions without mentioning visuals. 

Among the 228 comments with concrete visual details, 75\% (171/228) were written in the format of description, such as \code{0:09 This shows a sunny, outdoor open area with trees, paths, and buildings in the background. There is a group of students standing talking to each other and we see two students walking together on the path.}\footnote{\url{https://www.youtube.com/watch?v=WzJc75S-LzY}}, providing a detailed and objective description of a video scene without narrations.
The remaining 25\% (57/228) were in the format of sharing feelings but making an additional effort to make their comments more accessible by clarifying the visuals, such as \code{3:31 Oh my! There's one white fox in all of the reddish ones!}\footnote{\url{https://www.youtube.com/watch?v=nrSdiqp0Dpk}}, 
and \code{2:07 The cat is so cute chewing on the brush oh my gosh}\footnote{\url{https://www.youtube.com/watch?v=ORtlZG_RU1s}}, providing more visual contexts when expressing opinions with adjectives and interjections.
These comments provide descriptive details to scenes (78/228), concepts (52/228), actions (44/228), objects (30/228), animations (12/228), and texts (12/228) shown in the videos.

3\% (8/257) of the total comments only mentioned visual entities but didn't covered any concrete descriptive information, such as \code{3:21 various images to accompany the discussion}, and \code{4:54 the cells look very cute}.\footnote{\url{https://www.youtube.com/watch?v=qPix_X-9t7E}}
The former one indicated an attempt from the user trying to mention visuals but failing to provide informative details, and the latter one indicated a sharing of opinion from the user without making efforts in providing more descriptive details.
There also existed exceptions: sometimes mentioning the visual object was enough since it is not necessary to describe how the object looks in detail if it was not directly related to the video theme, or the detail was already covered in the audio track and can be redundant if described again.
Additionally, 8\% (21/257) of the total comments purely expressed participants' opinions without mentioning visuals, such as \code{Seattle is a beautiful city}.\footnote{\url{https://www.youtube.com/watch?v=WzJc75S-LzY}}

\subsection{Understanding Viewer Experience of Descriptive Commenting}
While this study was not intended as a formal evaluation of feature efficacy, participants’ experiences shed light on what makes descriptive commenting approachable, and what remains challenging.

\begin{figure*}[h]
  \centering
  \includegraphics[width=1.0\linewidth]{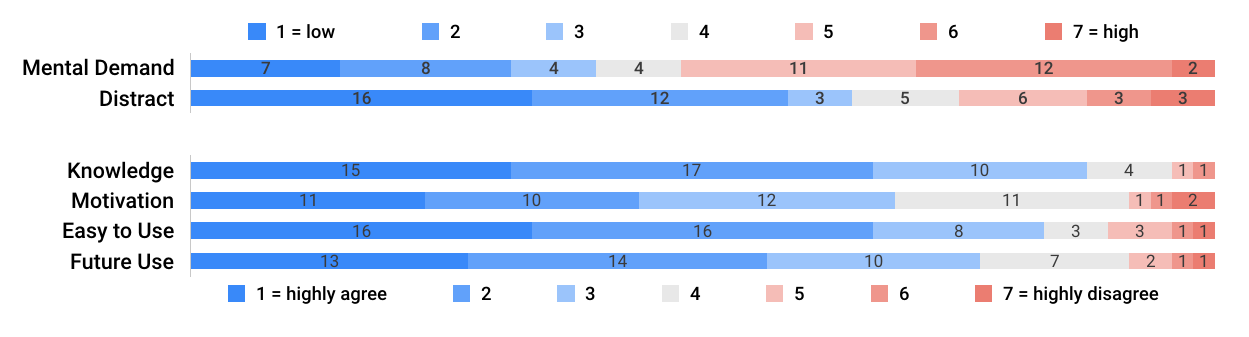}
  \caption{Participants reported ratings from 1 to 7 on their experience of descriptive commenting, including the following statements: 1) How mentally demanding was writing accessible comments? (1 = low, 7 = high) 2) How distracting of writing accessible comments to the video watching experience? (1 = low, 7 = high) 3) After using this tool, I understand the concept of video accessibility. (1 = highly agree, 7 = high disagree) 4) This tool motivates me to contribute to the accessibility of online videos. 5) This tool makes it easier for me when writing accessible comments. 6) If these features are built in Youtube, I will use this tool to make my comments more accessible when watching videos in the future.
  }
  \label{fig:sighted_rating}
\end{figure*}

\subsubsection{Perceived Value and Ease of Use}
Most participants expressed interest in using CoSight in the future (37/48). The tool empowered users to understand the concept of accessibility (42/48) and motivated them to contribute (33/48). Participants recognized CoSight’s value in promoting awareness of accessibility, noting that \textit{``it can help BLV people in ways that people don't normally think about''} and \textit{``it creates a positive environment that promotes collaboration among the audience''}. 

Most participants (40/48) agreed that CoSight makes contributing to accessibility easier, with some describing the interaction flow as \textit{``quite intuitive and simple''}, and the interactive features \textit{``fit smoothly with the YouTube video page.''}
Compared with providing a description in comments, modifying a comment to make it more accessible is much easier to do, as participants shared: \textit{``I like that there are just basic tweaks to ordinary comments to make them accessible. Like, if commenting on a particular frame, begin with a time stamp and avoid vague language or ambiguous pronouns.''} (Figure~\ref{fig:interaction_flow}F) and \textit{``It is an easier starting point for someone without much knowledge on accessibility and this would make more sighted people want to contribute.''}

\subsubsection{Interpretations of Individual Features}
Specifically, 10 participants mentioned the color labels (Figure~\ref{fig:interaction_flow}A) are helpful as an overview, as shared: \textit{``it helps sighted people realize where the video needs to be accessible.''}
7 participants mentioned the just-in-time signal (Figure~\ref{fig:interaction_flow}B) and pop-up hints (Figure~\ref{fig:interaction_flow}C) were helpful in reminding them of when and what they can describe;
5 participants highlighted that the shortcut of clicking on accessibility icon reduced their workload by automatically jumping to comment function and inputting the timestamp (Figure~\ref{fig:interaction_flow}D).
Showing related comments written by others (Figure~\ref{fig:interaction_flow}E) is mentioned as helpful 
as it saves their efforts for valuable contributions by \textit{``preventing writing redundant comments''}.
\subsubsection{Barriers to Contribution}
Some participants (12/48) reported feeling distracted from watching videos, with 4 especially mentioning when they had to manually rewind to make sure they wrote an informative comment for a video segment containing various visual details.
To optimize the flow, they suggested to support adding a \textit{`bookmark'} or \textit{`shortcut'} for one segment to revisit and rewind conveniently when they hoped to finish watching the video first.


Although participants generally found CoSight easy to use, 25 participants still reported that writing accessible comments felt mentally demanding, which stemmed from the nature of the task.
Some mentioned the mental efforts came from the mismatch between typical commenting motivation (which focus on sharing opinions) and the new expectation to include descriptive visuals details. As one participant reflected, \textit{``I don't normally write comments with descriptions, so this was a bit mentally taxing.''}
Others described the onboarding process of learning how to write accessible comments as mental demanding. However, they also agreed that this one-time learning step has long-term benefits, as one participant shared, \textit{``it was somewhat intimidating when not familiar with the concept at the beginning but much better when I get used to it, and I will be able to contribute better comments for blind people.''}


\subsection{Understanding Accessibility Values of Descriptive Comments}




To explore how viewer-contributed comments might offer value to BLV audiences, we interviewed both BLV individuals and professional AD writers, not to compare comments to professional descriptions, but to understand their potential as an informal, viewer-generated layer of accessibility.

\subsubsection{Descriptive Comments Are Not Equivalent to AD, but Can Offer Complementary Value}

All AD writers agreed that most of collected comments cannot compare with the quality of professional AD authoring,
as they mentioned two common problems of novice writers without training, either \textit{``including not enough details''} or \textit{``over-describing''}, also applied to these comments.
For example, when writing for a cat bathing scene, a collected comments just briefly touch on \code{the man is rubbing the cat now} while W1 demonstrated that she would provide more descriptions on actions such as \code{hands rotate around to massage and move up and down the cats body}.\footnote{\url{https://www.youtube.com/watch?v=ORtlZG_RU1s}}

However, AD writers emphasized that viewer comments are not intended to compete with the quality of professional AD, but could offer complementary value.
\begin{quote}
    \textit{I always spend a lot of time revising my descriptions, given the nature of such comments are written quickly just at the time of watching, at least it works well on providing more visual details.} (W1)
\end{quote}
\begin{quote}
    \textit{Although comments are not doing it equivalently as AD, it doesn't mean that it worth nothing. These comments can still promote video accessibility.} (W2)
\end{quote}

While AD professional writers have more stringent standards on the quality of descriptions due to their training experiences, BLV participants gave credit to comments providing visual details unknown to them.
When sharing their prior experience, all BLV users mentioned the comments are often very subjective and don't offer detailed information for understanding the visuals.
\begin{quote}
    \textit{It's usually people disagreeing or agreeing something. Or it'll be something like `Hey, did you see the thing at 2:55? That was really cool!' Well, that's not helpful without telling what exactly they're saying.} (U1)
\end{quote}

After watching videos with CoSight, BLV participants agreed that most comments with timestamps and concrete visual details were helpful. 
For example, a comment \textit{`1:29 Green algae swirls in the wake created by a tiger swimming across the screen.'}\footnote{\url{https://www.youtube.com/watch?v=FK3dav4bA4s}} describing a dynamic scene without speech, was regarded as helpful by U3 since it provided visual information that she would not have had.
Still, comments that purely expressing opinions without visual information were distracting to U1, like one comment of the same video, saying \textit{`It is crazy how Tigers prefer to be in the water contrary to most cats.'}.
\begin{quote}
    \textit{The majority of the comments were pretty helpful as far as being descriptive. But I don't want to make a blanket statement, say, `definitely all helpful'. I would say no to opinionated comments.} (U1)
\end{quote}



\subsubsection{Comments vs Descriptions: Subjective Opinions As Double-Edged Sword}
BLV participants' attitudes toward the subjective comments differ.
While U1 and U3 tend to not read comments with any subjective opinions when looking for descriptive information, U2 and U4 accepted reading descriptive comments with personal opinions involved, which can be fun and help them better understanding of video contents.
\begin{quote}
    \textit{Comments definitely helped with the context. Obviously, they can be very subjective, but it helps give me an idea like about little nuances that I would otherwise miss.} (U2)
\end{quote}
Occasionally, comments with opinions can be useful when it pointed out some errors in the visual frame, such as \code{1:10 Every part of the tongue detects sweetness. There's an inaccurate tongue map diagram.}\footnote{\url{https://www.youtube.com/watch?v=lEXBxijQREo}} 
Meanwhile, participants understood it is within expectations that everyone writes comments in their own language style and not everyone can write a professional description.
\begin{quote}
    \textit{You don't have to be a professional describer. It can be beneficial when humans helping out humans by telling me a few things that I'm missing. If it's not perfect, it is okay, we're still getting the info.} (U4)
\end{quote}
Interestingly, echoing with the mismatch between commenting and describing from sighted viewers, all BLV participants mentioned \code{nomenclature}, the wording used in the interface, can be important to nudge the sighted audience to share more objective descriptions than subjective feelings.
\begin{quote}
    \textit{The term `comment' may naturally let other people think 'Oh, they want my opinion of it' but the term description may let people feel 'you are asking me to relay information to you'.} (U1)
\end{quote}
All AD writers mentioned that there exists a grey area for subjective and objective contents in comments, which can be hard for general audience to distinguish and decide what to include and what to remove. While personal opinions are definitely not accepted in professional AD, they feel a mix of opinions and descriptions is acceptable in comments, as long as BLV people are aware of that these comments are not written by professionals.
\begin{quote}
    \textit{Different from audio descriptions, I think comments are just fun! Being totally objective will lose the fun. The vibes and emotions are also important information to convey.} (W2)
\end{quote}

\subsubsection{Social Benefits and Concerns}
BLV participants commented CoSight as \userquote{1}{a really dynamite service}, \userquote{2}{a neat tool}, and \userquote{3}{a good middle-ground solution of raising awareness}.
Different from YouDescribe, a separate crowdsourcing platform that volunteers add audio descriptions based on requests from visually impaired people, the spirit of CoSight is to take proactive actions as a sighted YouTube user without expertise in accessibility.
\begin{quote}
    \textit{I think it could be very useful to to blind audience. This reminds me of YouDescribe. But this [CoSight] is baked into YouTube in a more primitive way.} (U1)
\end{quote}
\begin{quote}
    \textit{I would definitely enjoy this quite a lot because of the intimidating way it approached to improve awareness...the idea is like, we're not throwing out like 508 compliance, ADA standard, it's not like scary and a ton of extra work. Once people understand this, it's not super complicated to contribute.} (U3)
\end{quote}
AD writers also see the value of CoSight not only in improving accessibility of specific videos for BLV people, but also promoting awareness of contributing to accessibility among sighted people, which has long-term social benefits.
\begin{quote}
    \textit{It may not be most accurate or most beneficial in providing descriptions from the standard of AD, but it does raise consciousness and awareness to engage more people in the society in video accessibility.} (W4)
\end{quote}

The main concerns mentioned by participants are around bias and abuse.
\wquote{3}{We all have a description bias. If I say something is large, my large might be your small. It is necessary to educate people to try to keep their biases out when writing descriptions.}
While in the current development, CoSight presented human-authoring comments without reviewing, AD writers worry about deploying it on a large scale.
\userquote{1}{Now we are testing it in a really controlled scenario. But when you put it in the wild, it's going to be unordered chaos. People may abuse the system and data.} 

\subsubsection{Reflections on Results}
These interviews suggest that while viewer-generated comments cannot match the rigor of professional AD, but can offer complementary value when grounded in concrete visual details. Participants appreciated descriptive elements that conveyed visual context, but concerns around subjectivity, bias, and quality control highlight the importance of positioning such contributions as a complement, not a replacement, for formal accessibility. 
Using CoSight as a design probe, this exploratory study represents an initial step toward understanding the feasibility of inviting casual viewer participation and the tensions that community-driven accessibility efforts can introduce. Keeping the door open to such contributions is not enough. Meaningful integration into accessibility must be supported by responsible curation, clear scaffolding, ongoing reflection, and transparency about their role. Without this care, casual contributions risk creating a false sense of completeness while falling short of access needs.
\section{Discussion}
This work explores the feasibility and potential of engaging everyday media consumers in supporting video accessibility through descriptive comments. Building on insights from a survey of internet users, and an exploratory study using CoSight with sighted YouTube users, BLV people, and professional AD writers, we shift the focus to video viewers, those who typically consume rather than produce content. While mindsets and perceived responsibility vary across individuals and contexts, our findings suggest that aligning contribution opportunities with users’ existing viewing behaviors can foster new forms of participation. In this section, we reflect on lessons learned, limitations of this work, and opportunities for future work.

\subsection{Scaffolding Participation While Safeguarding Accessibility Integrity}
Most sighted participants found CoSight intuitive and easy to use, especially with features that reduce the effort required to author a comment with timestamps and visual details.
However, the challenge lay not in using the tool itself, but in adjusting to a new mental model: from opinion-sharing to visual description. This cognitive shift, particularly for first-time users unfamiliar with accessibility practices, can be mental demanding.
In our study, participants first spent time on reading through the instructions on writing accessible comments.
In real-world scenarios, it is unlikely that casual viewers will take the time to read instructions before watching a video or commenting. This suggests the need for further refinement in how accessibility knowledge is gradually integrated into the viewing experience, rather than delivered all at once.
Scaffolding is a frequently used strategy to facilitate learning by enabling learners to accomplish tasks that would be beyond their capacity if unassisted \cite{park2022study}.
Instead of requiring users to read instructions upfront, pop-up cards could serve as educational tools during the video, offering examples of high-quality descriptions. This would allow viewers to learn about accessibility as they watch, gradually developing their understanding. 

Scaffolding alone is not enough. A key concern from interviews was the risk of false completeness — where casual contributions might be treated as sufficient alternatives, sidelining the need for professional accessibility. To avoid this, viewer-contributed content must be clearly labeled and responsibly curated. Future work could explore lightweight moderation pipelines such as peer review or community vetting to elevate useful comments and filter out harm. Looking ahead, we envision CoSight as a tiered learning platform: starting with small, low-effort contributions and gradually supporting more experienced and verified users in crafting high-quality visual descriptions. This would allow us to maintain an inclusive on-ramp while safeguarding accessibility integrity.


\subsection{Incentivizing People with Rewarding}
Most sighted participants reported being motivated to contribute to video accessibility using CoSight, also revealing potential opportunities to make the process more engaging and enjoyable through rewarding strategies.
Gamification is defined in HCI as the use of game design elements in non-game contexts~\cite{10.1145/2181037.2181040}, which applies game elements such as challenges, levels, avatars, points, achievements, stories, and leaderboards into applications which their main purpose goes beyond pure entertainment. 
For instance, when a viewer submits an accessible comment, a celebratory animation could be shown to reinforce the achievement. CoSight already tracks a user's contribution history, allowing the introduction of a leaderboard to highlight active contributors for specific YouTube channels. Milestones such as contributing 10, 100, or 1,000 accessible comments could be recognized with user titles or badges.

In addition, sighted viewers could receive upvotes or feedback from BLV viewers, providing a sense of accomplishment and social influence. The platform could also visualize the impact of contributions by displaying the number of people helped through accessible comments.
As an alternative incentive, platforms could allow users to write an accessible comment in exchange for skipping ads. Rather than a one-size-fits-all approach, offering variable designs, triggers, and rewards could help engage a more diverse group of contributors.

\subsection{Potential and Concerns of Involving AI in the Loop}
Researchers have explored different approaches of generating descriptions of visual media, including fully human-authored (\eg \cite{10.1145/2049536.2049540}), fully automated (\eg \cite{devlin2015language}) and hybrid techniques of computer vision algorithms co-authoring with humans (\eg \cite{10.1145/3313831.3376728}).
Most efforts in mainstream solutions have concentrated on achieving acceptable results based on the visual media itself, but it is becoming evident that researchers need to consider different contexts into content generation\cite{10.1145/3441852.3471233}.
Vision language models can be applied to generate descriptions with multi-modal meta-data, including video itself, captions, human-authoring comments.
Meanwhile, involving AI for summarizing the comments may lose the creativity and diversity of human-authored contents, also aggravate prejudice if a similar expression is used in multiple comments even it is harmful.
Although there is potential to reduce the authoring efforts by supporting human to modify a generate result, researchers should also be cautious when involving AI in nudging for accessibility, as prior research on alt text \cite{10.1145/3441852.3471207} authoring has revealed that representing AI-generated info to assist content authoring may decrease the quality. 
Therefore, in this work, we deliberately avoid AI interference in content authoring but leverage its strengths in identifying and explaining inaccessible segments.
While some participants contributed comments for video segments not flagged as inaccessible by the AI, most followed the AI’s indicators. This highlights the importance of preserving user autonomy and avoiding over-reliance on AI suggestions in future designs.

Importantly, viewer-contributed comments and AI-generated descriptions offer different strengths. AI output can provide consistency, coverage, and speed at scale, but it often lacks the emotional nuance, contextual awareness, and diversity of expression that characterize human-authored comments. In contrast, viewer comments, though informal and variable in quality, can reflect real-world interpretations, capture humor or sentiment, and provide a more socially grounded perspective. Rather than positioning one approach as a replacement for the other, we suggest future opportunities for hybrid solutions fostering cooperation between AI and humans. Future design could engage AI in rephrasing human-authored comments into a more curated audio description style, enhancing the overall quality without overshadowing the human contribution or pushing human write comments in description style.

\subsection{Values and Challenges of Deploying in the Wild}

Reflecting on YouTube’s history, the platform once allowed community captions, enabling viewers to add subtitles to videos. This feature was discontinued in 2020 due to low usage and abuse, underscoring the need to design solutions that guide users toward desired behaviors in tailored ways.
While we may address the algorithmic challenges of processing data with bias by leveraging technical solutions, policymakers and platforms should make a continued effort on developing socio-technical solutions to promote social awareness of accessibility and avoid abuse \cite{10.1145/3532106.3533514, 10.1145/3411764.3445321}. More broadly, the concept of engaging media consumers in contributing to accessibility can be extended to various online platforms. For instance, users could add more descriptive visual details in product reviews, comments for visual media like comics, or even share accessibility-related insights about sightseeing spots.

As an exploratory study, our work does not fully capture the nuances of viewer behavior in naturalistic, in-the-wild settings. Participants interacted with a curated set of videos under study conditions, and while the findings offer valuable insights, they stop short of revealing how viewers might engage with accessibility contributions in their everyday video-watching routines.
Future research could benefit from longer-term deployments that observe participants over time — comparing behaviors with and without just-in-time triggers, across videos that align with their personal interests and habits. Expanding the study to include a wider range of video genres and lengths would provide a deeper understanding of how content context affects accessibility contributions. Such efforts, however, would require both technical scalability and partnership with video platforms to enable real-time support and data integration.
Additionally, exploring the experiences of BLV viewers using comments as an accessibility resource when watching videos, and fostering interactions between BLV and sighted viewers, remains a valuable area for further study.
BLV people and sighted people can work together to co-create an accessible household~\cite{10.1145/2702123.2702511} and workspace ~\cite{10.1145/2700648.2809864}, suggesting there is value in exploring similar collaborations for online communities.





\section{Conclusion}
This work explores a new perspective in accessibility research: engaging everyday video viewers as potential contributors to online media accessibility. Through a survey, a design probe, and an exploratory study involving sighted viewers, BLV individuals, and professional AD writers, we examined the feasibility and value of descriptive commenting as a lightweight, socially embedded form of accessibility support. While viewer-authored comments cannot replace professional AD, our findings show that such triggered contributions can offer complementary value: adding visual grounding, emotional nuance, and a sense of shared experience to the viewing context. We believe accessibility should always center those who rely on it. At the same time, we see value in engaging a broader public to care, contribute, and learn as part of a more inclusive accessibility ecosystem.
We recognize the risks of low-quality or chaotic contributions, but also believe it’s important to keep the door open to casual participation, even if it doesn’t initially meet best-practice standards. Such participation must be supported through responsible curation, thoughtful scaffolding, ongoing reflection, and honest communication about its role in accessibility. We hope CoSight contributes to this conversation, and sparks continued efforts toward community-driven, inclusive participation in accessibility.

\begin{acks}
The authors thank all study participants who generously shared their time and experience for this work.
\end{acks}

\bibliographystyle{ACM-Reference-Format}
\bibliography{main}

\appendix
\section{Instructions on writing accessible comments with CoSight}

\subsection{Why your comments may help blind people}
Describing visual information can help blind and visually impaired people who cannot see the video adequately. Professionals add descriptions to the original audio tracks for films and some videos. It is not realistic to have every online video described by professionals or content creators, however, every ordinary audience like you has the power of contributing to the accessibility of the video you are watching, by making your comment more accessible by clarifying the visual, or writing a descriptive comment, especially for an inaccessible segment! When blind people are listening to the videos, we make a beep sound if helpful comments are available for a segment, then they can choose to pause the video, use screen readers to navigate, and read out the selected comments.\\

\subsection{Make your comment more accessible by clarifying the visual}
Add timestamp if you are talking about specific segments of the video; Avoid using ambiguous pronouns, such as it, that, or this, instead, use the full names or concepts you refer to. If you share feelings on visual content not directly covered in the audio track, describe the visual content. Here is an example of modifying a comment:

\begin{itemize}
    \item \textbf{Original}: I like it when he walked in. The scene was so cool!
    \item \textbf{More accessible}: 5:26 I like it when the vlogger’s friend walked in. The scene of all the lights turning on behind him was so cool!
\end{itemize}

Here are more examples of comments with clarified visuals:
\begin{itemize}
    \item 1:29 WAIT A MINUTE! Taste perception is not localized like that in the brain. Different tastes are not processed in different parts of the cerebral cortex as the illustration showed.
    \item 1:42 thanks, now my legs are going to be itchy for the rest of the video because of the little animation of the spider moving on the skin
\end{itemize}

\subsection{Write a description, especially for an inaccessible segment}
Visual elements that are important to understand what the video is communicating:
\begin{itemize}
    \item Objects (e.g. ingredients needed for a recipe); 
    \item Actions (e.g. a step in a tutorial video); 
    \item Scenes (e.g. a person studying); 
    \item Animations (e.g., an arrow showing the direction of force); 
    \item Texts (e.g., additional info not mentioned in the audio).
\end{itemize}

Don’t need to describe every detail: Apparent from the audio. (e.g. host speaking to camera); Already mentioned/described in the audio.

The system may not identify every segment that deserves a description, feel free to add comments to where you think is meaningful!

Here are some examples of descriptive comments:

\begin{itemize}
    \item 0:28 I learned a little trick here: reverse the pan and use the back of the pan to form the shape of foil
    \item 0:45 She showed a white pan and a black pan. If you use a black pan, remember to reduce the temperature of your oven by 25 degrees
    \item 1:39 It seems that she used at least three eggs
\end{itemize}

\section{Explaining inaccessible segments and Instructions on What to Describe}
The reasons of inaccessibility followed the definitions from prior work \cite{10.1145/3411764.3445233}.

\begin{itemize}
    \item  \textbf{R1 Contain no-speech segment}: clarify the visual contents
    \item \textbf{R2 Not informative language}: clarify the concrete details of visual contents
    \item \textbf{R3 Frequent shot changes}: clarify the changing scenes
    \item \textbf{R4 Numerous visual entities}: clarify any important visual entities
    \item \textbf{R5 Specific visual details not covered by speech}: clarify any missing visual details
    \item \textbf{R6 On-screen text not in speech}: clarify the important on-screen text
    \item \textbf{R7 Reference words without explanations}: clarify it/this/that words in the speech
\end{itemize}

\section{Comments Collected by CoSight in Study with Sighted Audience}


Table 1 listed example comments written by participants using CoSight, illustrating both types of contributions.

For instance, \textit{``1:42 Seeing an image of a spider crawling on a person’s leg hair makes me cringe!''} demonstrates how a reaction-based comment being refined to be more accessible by grounding the response in specific visual content.

In contrast, \textit{``8:60 neurons are shown going down the nerves in the image of a dark body from the top to the leg''} reflects an attempt to describe visual information not conveyed through audio — aligning with the intention of audio description, though in a casual form.

Video name, ID, and screenshots of YouTube videos used in the study are shown in Figure 7.

\begin{table*}[h]
\caption{Some example descriptive comments sighted participants wrote in the study.}
\label{tab:my-table}
\begin{tabular}{l|l}
\textbf{Video Title}                                                                                                        & \textbf{Some Descriptive Comments Written with CoSight}                                                                                                                                                                                                                                                               \\ \hline
\begin{tabular}[c]{@{}l@{}}The Nervous System, Part 1: \\ Crash Course Anatomy \& Physiology \#8\end{tabular}               & \begin{tabular}[c]{@{}l@{}}8:60 neurons are shown going down the nerves in the image of a dark body from \\the top to the leg \\ 1:42 Seeing an image of a spider crawling on a person's leg hair makes me cringe!\end{tabular}                                                                                                                                     \\ \hline
\begin{tabular}[c]{@{}l@{}}How to Bathe your Cat that Hates \\ Water (6 Step Tutorial) | The Cat Butler\end{tabular}        & \begin{tabular}[c]{@{}l@{}}0:26 a man pours water onto a wet cat inside the bath tub from a pitcher \\ 5:01 the man is trying to dry off his cat with a towel but the cat keeps trying to get away \end{tabular}                                                                                                                                                                                                       \\ \hline
\begin{tabular}[c]{@{}l@{}}Fox Village in Japan: The Fluffiest Place \\ on Earth! \end{tabular} & \begin{tabular}[c]{@{}l@{}}0:17 There's about 30 rabbits surrounding that woman! Too cute! They look so fluffy and \\ soft and clean. \\  2:13 The narrator is wearing a faux fur lined coat, and she made a note in the video \\ letting us know it's fake!\end{tabular}                                                                                                                              \\ \hline
How sugar affects the brain - Nicole Avena                                                                                  & \begin{tabular}[c]{@{}l@{}}At 1:12 they are showing a drawing of a tongue and highlighting different parts to show \\ where on your tongue different tastes are felt. \\ 1:51 It's raining candy! The words are shaped into a slot machine that has flashing lights! \\ They really went all out on this one. The words read "Should I do that again? and then in big \\ letters, 'GO FOR IT!" \end{tabular}                                                                                              \\ \hline
Tigers 101 | National Geographic                                                                                            & \begin{tabular}[c]{@{}l@{}}0:56 The introductory scenes are of select shots of tigers in the wild, beautiful creatures, \\ and one even looks at the camera. \\  1:29 just seeing the tiger swimming through the green algae filled swamp makes me feel slimy! \end{tabular}                                                                                                       \\ \hline
How to Plant Potatoes! Garden Answer                                                                                        & \begin{tabular}[c]{@{}l@{}}0:29 She is sitting on the ground by a raised planter. Beside her is a tray of potatoes, and she is \\ touching the dark, loamy soil with her bare hand. Behind her is a pitchfork and the bag the \\organic gardening soil came from. \\  1:59 she  is making some evenly trenched rows in the dirt to prepare the dirt in the planter box \\ for the potatoes. very nice and even rows of dirt with the trenches dug about 6 to 10 inches deep.\end{tabular} \\ \hline
UW Campus Tour: Hear It from a Husky                                                                                        & \begin{tabular}[c]{@{}l@{}}1:16 The red square is such a large open area between multiple buildings, \\ layed with burgandy colored bricks. Very stunning. \\ 0:23 The campus looks so peaceful being surrounded by so many cherry trees. \end{tabular}                                                                                                         \\ \hline
\begin{tabular}[c]{@{}l@{}}Experiment! How Does An Owl Fly So\\  Silently? | Super Powered Owls | BBC\end{tabular}          & \begin{tabular}[c]{@{}l@{}}2:05 The owl appears to have red eyes in a zoomed view. The red eyes and white fur \\ give me the impression this is an albino variant. \\  1:17 I love the smiles on the faces of the individuals recording the birds in flight, \\the amazement. :)\end{tabular}                                                                                                
\end{tabular}
\end{table*}

\begin{figure*}[h]
  \centering
  \includegraphics[width=1.0\linewidth]{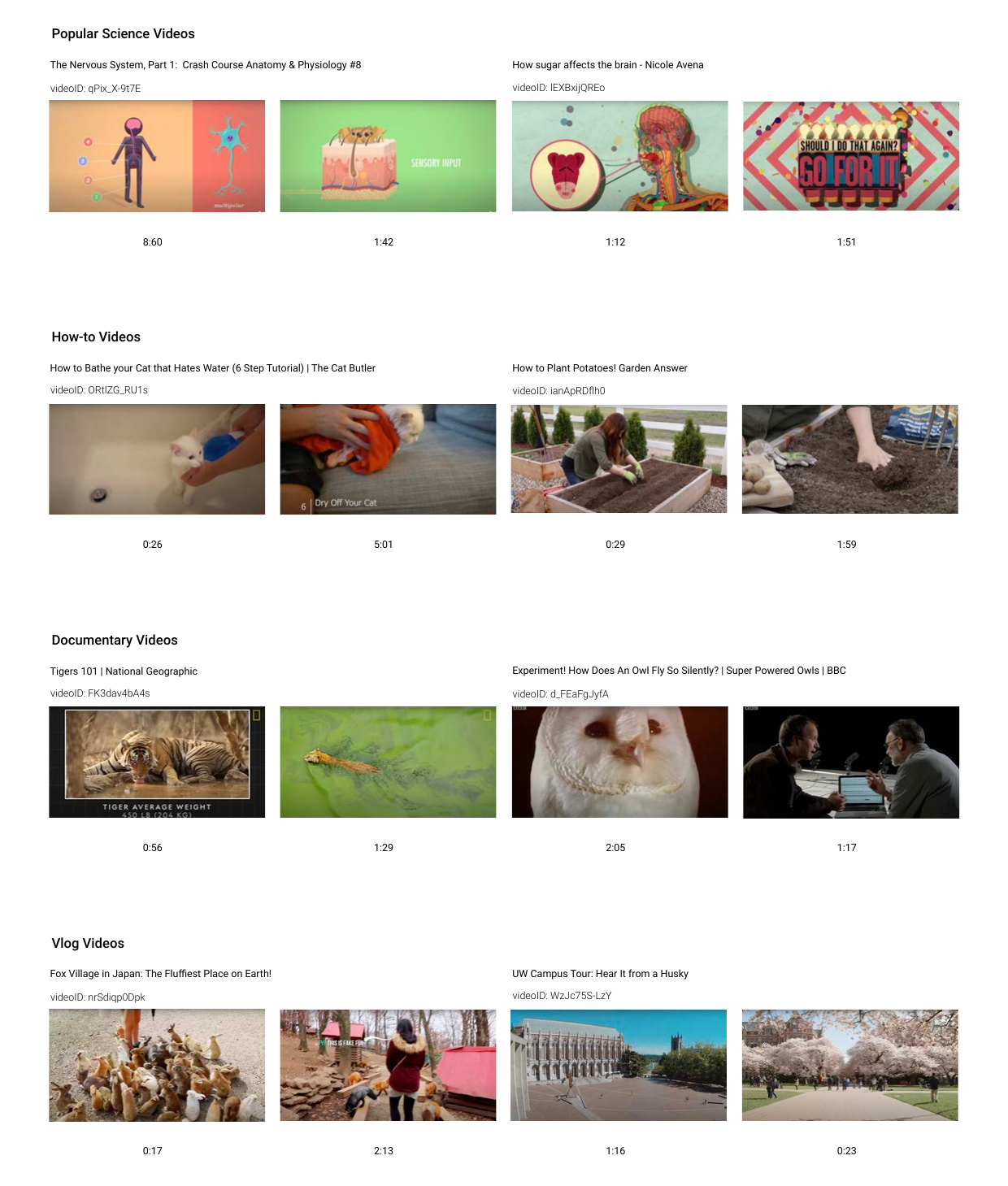}
  \caption{YouTube videos used in the study. Screenshots correspond to the timestamps of participant comments listed in Table 1.}
  \label{fig:appendix}
\end{figure*}

\section{Demographic Info}
Table 2 listed demographic info of blind participants in the study.

\begin{table*}[h]
\vspace{1em}
   \caption{Demographics of blind participants.}
   \label{tab:demo}
   \begin{tabular}{ccccc}
     \toprule
     ID & Age/Gender & Visually Impairment & Audio Description Experience\\
     \midrule
     \ U1 & 37/M & Totally Blind & Use it in daily life \\
     \ U2 & 28/F & Totally Blind & Use it in daily life \\
     \ U3 & 34/F & Low Vision & Use it in daily life and also work on Professional training of AD writers \\
     \ U4 & 41/M & Totally Blind & Use it in daily life \\
   \bottomrule
    \end{tabular}
\end{table*}

\end{document}